\documentclass[twocolumn,nofootinbib,longbibliography]{revtex4-1}

\usepackage[english]{babel}
\usepackage{appendix}
\usepackage{natbib}

\usepackage{amsmath}
\usepackage{mathrsfs}
\usepackage{amssymb}
\usepackage{amsfonts}
\usepackage{dsfont}
\usepackage{IEEEtrantools}
\usepackage{bm}
\usepackage{amsthm}

\usepackage{hyperref}
\usepackage[inline]{enumitem}

\makeatletter
\usepackage[threshold=1, autopunct, autostyle]{csquotes}
\makeatother

\renewcommand{\H}{\mathcal{H}}
\newcommand{\ket}[1]{|#1 \rangle}

\newcommand{\proj}[1]{|#1\rangle\langle #1|}
\newcommand{\da}{\downarrow}
\newcommand{\ua}{\uparrow}

\DeclareMathOperator{\tr}{Tr}

\usepackage{pgfplots}
\usepackage{tikz}
\usetikzlibrary{shapes,calc,positioning,arrows.meta}

\pgfdeclarelayer{bg}    
\pgfsetlayers{bg,main}

\tikzset{colorbox/.style={thick, rounded corners=2pt, text height=1.7ex,text depth=.25ex, draw=#1!70!black, fill=#1!30}}
\tikzset{colorboxS/.style={thick, rounded corners=2pt, text height=1.2ex,text depth=.25ex, draw=#1!70!black, fill=#1!30, font=\footnotesize}}
\tikzset{colorboxXS/.style={thick, rounded corners=2pt, text height=0.7ex,text depth=.10ex, draw=#1!70!black, fill=#1!30, font=\tiny}}
\tikzset{roundedbox/.style={thick, rounded corners=2pt, text height=1.7ex,text depth=.25ex, draw=black}}
\tikzset{CBedgy/.style={thick, text height=1.7ex,text depth=.25ex, draw=#1!70!black, fill=#1!30, font=\ttfamily}}
\tikzset{colorelement/.style={thick, rounded corners=2pt, draw=#1!70!black, fill=#1!30}}
\tikzset{quote/.style={thick, draw=black!70, rounded corners=2pt, font=\footnotesize, text width=#1, text opacity=1, opacity=.8, fill=white},
  quote/.default=4.5cm}
\tikzset{quoteS/.style={thick, draw=black!70, rounded corners=2pt, font=\scriptsize, text width=#1, text opacity=1, opacity=.8, fill=white},
  quoteS/.default=4.5cm}
\tikzset{quoteXS/.style={thick, draw=black!70, rounded corners=2pt, font=\tiny, text width=#1, text opacity=1, opacity=.8, fill=white},
  quoteXS/.default=4.5cm}
\tikzset{quoteNB/.style={font=\footnotesize, text width=#1, text opacity=1, opacity=.8, fill=white},
  quoteNB/.default=4.5cm}
\tikzset{comment/.style={thick, draw=black!70, rounded corners=2pt, font=\scriptsize\itshape, text width=#1, text opacity=1, opacity=.8, fill=white},
  comment/.default=4.5cm}
\tikzset{commentS/.style={thick, draw=black!70, rounded corners=2pt, font=\tiny\itshape, text width=#1, text opacity=1, opacity=.8, fill=white},
  commentS/.default=4.5cm}
\tikzset{commentVW/.style={thick, draw=black!70, rounded corners=2pt, font=\scriptsize\itshape, text opacity=1, opacity=.8, fill=white}}
\tikzset{commentSVW/.style={thick, text height=0.7ex,text depth=.10ex, draw=black!70, rounded corners=2pt, font=\tiny\itshape, text opacity=1, opacity=.8, fill=white}}
\tikzset{conn/.style={thick, shorten <=#1, shorten >=#1}}
\tikzset{tconn/.style={shorten <=#1, shorten >=#1}}
\tikzset{gconn/.style={thick, shorten <=#1, shorten >=#1, draw=gray!80}}
\tikzset{arr_node/.style={pos=0.5,above,font=\scriptsize, sloped}}
\tikzset{ctag/.style={thick, dashed, rounded corners=2pt, text height=1.7ex,text depth=.25ex, draw=#1!70!black, fill=#1!30, font=\ttfamily}}

\tikzset{ist/.style={thick, shorten <=4pt, shorten >=4pt, arrows = {-Bracket[reversed,round]}}}
\tikzset{istgleich/.style={thick, shorten <=4pt, shorten >=4pt, arrows = {Bracket[reversed,round]-Bracket[reversed,round]}}}
\tikzset{nicht/.style={thick, dashed, shorten <=4pt, shorten >=4pt, arrows = {Bracket[round]-}}}

\definecolor{yorange}{HTML}{ff8c00}

\newcommand\blArrow{}
\def\blArrow[#1](#2);
  { \draw[#1] (#2) -- ++(.9,0) -- ++(-.9,-5) -- cycle; }
\newcommand\stdBlArrow{}
\def\stdBlArrow(#1)
  { \blArrow[line width=2pt, draw=red!70!black, fill=red!40, rounded corners=1pt](#1) }

\newcommand\redArrow{}
\def\redArrow(#1);
  { \draw[thick] ($(#1)+(-.3,.1)$) -- ($(#1)+(0,-.1)$) -- ($(#1)+(.3,.1)$);}
\newcommand\fSTS{}
\def\fSTS(#1);
  { \draw[fill=red!40, thick, draw=red!60!black] (#1) circle (.5);
    \draw[->] ($(#1)+(-.05,-.05)$) to ++(0,.4);
    \draw[->] ($(#1)+(-.05,-.05)$) to ++(.4,0);
    \draw[->] ($(#1)+(-.05,-.05)$) to ++(-.2,-.2);}
\newcommand\reduction{}
\def\reduction(#1);
  { \draw[fill=cyan!40, thick, draw=cyan!60!black] (#1) circle (.5);
    \draw[fill=black] ($(#1) + (-.2, .25)$) to ++(0.4,0) to ++(-.3,-.6) to cycle;}
\newcommand\customLegend{}
\def\customLegend(#1);
  {\fSTS(#1);
   \node[right of=#1] {fixed space-time structure};
   \draw[rounded corners=2pt] ($(#1)+(-.5,-.5)$) rectangle ($(#1)+(4, -4)$);}

\makeatletter
\newcommand*{\blitzset}{\pgfqkeys{/blitz}}
\blitzset{
  height/.initial=4cm,
  width/.initial=2cm,
  breadth/.initial=.3cm,
  ratio/.initial=.4,
}

\pgfdeclareshape{blitz}
{
  \savedanchor\centerpoint{
    \pgf@x = .5\wd\pgfnodeparttextbox
    \pgf@y = .5\ht\pgfnodeparttextbox
  }
  \anchor{center}{\centerpoint}
  \anchor{north}{\centerpoint\pgfmathsetlength\pgf@ya{\pgfkeysvalueof{/blitz/ratio}*\pgfkeysvalueof{/blitz/height}}\advance \pgf@y by \pgf@ya}
  \anchor{south}{\centerpoint\pgfmathsetlength\pgf@ya{-(1-\pgfkeysvalueof{/blitz/ratio})*\pgfkeysvalueof{/blitz/height}}\advance \pgf@y by \pgf@ya}
  \anchor{east}{\centerpoint\pgfmathsetlength\pgf@ya{\pgfkeysvalueof{/blitz/breadth}}\advance \pgf@y by \pgf@ya\pgfmathsetmacro{\alpha}{atan(2*\pgfkeysvalueof{/blitz/ratio}*\pgfkeysvalueof{/blitz/height}/\pgfkeysvalueof{/blitz/width})}\pgfmathsetlength\pgf@xa{.5*\pgfkeysvalueof{/blitz/width}+\pgfkeysvalueof{/blitz/breadth}/tan(\alpha/2)}\advance\pgf@x by \pgf@xa}
  \anchor{west}{\centerpoint\pgfmathsetlength\pgf@ya{\pgfkeysvalueof{/blitz/breadth}}\advance \pgf@y by -\pgf@ya\pgfmathsetmacro{\alpha}{atan(2*\pgfkeysvalueof{/blitz/ratio}*\pgfkeysvalueof{/blitz/height}/\pgfkeysvalueof{/blitz/width})}\pgfmathsetlength\pgf@xa{.5*\pgfkeysvalueof{/blitz/width}+\pgfkeysvalueof{/blitz/breadth}/tan(\alpha/2)}\advance\pgf@x by -\pgf@xa}

  \backgroundpath{
    \centerpoint \pgf@xa=\pgf@x \pgf@ya=\pgf@y
    \pgfmathsetmacro{\alpha}{atan(2*\pgfkeysvalueof{/blitz/ratio}*\pgfkeysvalueof{/blitz/height}/\pgfkeysvalueof{/blitz/width})}
    \pgfmathsetlength\pgf@xb{.5*\pgfkeysvalueof{/blitz/width}-\pgfkeysvalueof{/blitz/breadth}/tan(\alpha/2)}
    \pgfmathsetlength\pgf@yb{\pgfkeysvalueof{/blitz/breadth}}
    \advance \pgf@xa by \pgf@xb
    \advance \pgf@ya by -\pgf@yb
    \pgfpathmoveto{\pgfpoint{\pgf@xa}{\pgf@ya}}
    \pgfmathsetlength\pgf@xb{\pgfkeysvalueof{/blitz/width}}
    \advance \pgf@xa by -\pgf@xb
    \pgfpathlineto{\pgfpoint{\pgf@xa}{\pgf@ya}}
    \pgfmathsetlength\pgf@yb{\pgfkeysvalueof{/blitz/breadth}+\pgfkeysvalueof{/blitz/ratio}*\pgfkeysvalueof{/blitz/height}}
    \pgfmathsetlength\pgf@xb{\pgf@yb*sin(90-\alpha)}
    \advance \pgf@xa by \pgf@xb
    \advance \pgf@ya by \pgf@yb
    \pgfpathlineto{\pgfpoint{\pgf@xa}{\pgf@ya}}
    \pgfmathsetlength\pgf@xb{2*\pgfkeysvalueof{/blitz/breadth}/cos(90-\alpha)}
    \advance \pgf@xa by \pgf@xb
    \pgfpathlineto{\pgfpoint{\pgf@xa}{\pgf@ya}}
    \pgfmathsetlength\pgf@yb{\pgfkeysvalueof{/blitz/ratio}*\pgfkeysvalueof{/blitz/height}-\pgfkeysvalueof{/blitz/breadth}}
    \pgfmathsetlength\pgf@xb{\pgf@yb*sin(90-\alpha)}
    \advance \pgf@xa by -\pgf@xb
    \advance \pgf@ya by -\pgf@yb
    \pgfpathlineto{\pgfpoint{\pgf@xa}{\pgf@ya}}
    \pgfmathsetlength\pgf@xb{\pgfkeysvalueof{/blitz/width}}
    \advance \pgf@xa by \pgf@xb
    \pgfpathlineto{\pgfpoint{\pgf@xa}{\pgf@ya}}
    \pgfmathsetlength\pgf@yb{(1-\pgfkeysvalueof{/blitz/ratio})*\pgfkeysvalueof{/blitz/height}+\pgfkeysvalueof{/blitz/breadth}}
    \pgfmathsetlength\pgf@xb{.5*\pgfkeysvalueof{/blitz/width}+\pgfkeysvalueof{/blitz/breadth}/tan(\alpha/2)}
    \advance \pgf@xa by -\pgf@xb
    \advance \pgf@ya by -\pgf@yb
    \pgfpathlineto{\pgfpoint{\pgf@xa}{\pgf@ya}}
    \pgfpathclose
  }
}

\newcommand*{\srefset}{\pgfqkeys{/sref}}
\srefset{
  height/.initial=.5cm,
  width/.initial=.5cm,
}

\pgfdeclareshape{sref}
{
  \savedanchor\centerpoint{
    \pgf@x = .5\wd\pgfnodeparttextbox
    \pgf@y = .5\ht\pgfnodeparttextbox
  }
  \anchor{center}{\centerpoint}
  \anchor{north}{\centerpoint\pgfmathsetlength\pgf@ya{.5*\pgfkeysvalueof{/sref/height}}\advance \pgf@y by \pgf@ya}
  \anchor{south}{\centerpoint\pgfmathsetlength\pgf@ya{-.5*\pgfkeysvalueof{/sref/height}}\advance \pgf@y by \pgf@ya}
  \anchor{east}{\centerpoint\pgfmathsetlength\pgf@xa{.5*\pgfkeysvalueof{/sref/width}}\advance \pgf@x by \pgf@xa}
  \anchor{west}{\centerpoint\pgfmathsetlength\pgf@xa{-.5*\pgfkeysvalueof{/sref/width}}\advance \pgf@x by \pgf@xa}

  \backgroundpath{
    \centerpoint \pgf@xa=\pgf@x \pgf@ya=\pgf@y
    \pgfmathsetlength\pgf@xb{.5*\pgfkeysvalueof{/sref/width}}
    \pgfmathsetlength\pgf@yb{.5*\pgfkeysvalueof{/sref/height}}
    \pgfmathsetlength\pgf@yc{.75*\pgfkeysvalueof{/sref/height}}
    \advance \pgf@xa by \pgf@xb
    \pgfpathmoveto{\pgfpointpolar{20}{\pgf@xa}}
    \pgfpatharc{20}{170}{\pgf@xb}
    \pgfpathlineto{\pgfpointpolar{140}{\pgf@yc}}
    \pgfpathmoveto{\pgfpointpolar{200}{\pgf@xa}}
    \pgfpatharc{200}{350}{\pgf@xb}
    \pgfpathlineto{\pgfpointpolar{320}{\pgf@yc}}
  }
}
\makeatother

\makeatletter
\@ifclassloaded{beamer}
  {}
  {

    \theoremstyle{remark}
    
    \theoremstyle{definition}

  }
\makeatother

\newif\ifradical
\radicaltrue
\newif\ifbeta
\betatrue
\newif\ifuncertain
\uncertaintrue
\newif\ifcomments
\commentstrue

\newcounter{CtrSprachspiel}

\makeatletter\begin{document}

\title{The Measurement Problem Is the ``Measurement'' Problem}

\author{Arne Hansen and Stefan Wolf}
\affiliation{Facolt\`a di Informatica, 
Universit\`a della Svizzera italiana, Via G. Buffi 13, 6900 Lugano, Switzerland}

\date{\today}

\begin{abstract}
  \vspace{15pt}
  \noindent
  The term ``measurement'' in quantum theory (as well as in other physical theories) is ambiguous:
It is used to describe both an \emph{experience}---e.g., an observation in an experiment---and an \emph{interaction} with the system under scrutiny.
If \emph{doing physics} is regarded as a creative activity to develop a meaningful description of the world, then one has to carefully discriminate between the two notions:
\emph{An observer's account of experience---consitutive to meaning---is hardly expressed exhaustively by the formal framework of an interaction within one particular theory.}
We develop a corresponding perspective onto central terms in quantum mechanics in general, and onto the measurement problem in particular.
   \vspace{25pt}
\end{abstract}

\maketitle

\section{Introduction}
\noindent
In the attempt to close quantum theory to the end of the observer---to incorporate an observer into the formal language of the theory---, one encounters problems such as the measurement problem~\cite{Maudlin95,BHW16}.
The Wigner's-friend experiment~\cite{wigner1963problem, deutsch1985quantum, Wigner1961} is a manifestation thereof:
\emph{Wigner} measures his friend who, in turn, measures another system.
The friend supposedly obtains a definite result as he ``performs a measurement'' while being, at the same time, in a superposition of states corresponding to two different results.
Wigner's friend is represented in the formal language of \emph{quantum} mechanics despite carrying some \emph{``classical''}~(\emph{distinguishable}) information.
Similarly, when one considers self-measurement scenarios, as in~\cite{Breuer1995}, the observer himself gets assigned a state, i.e., a symbol from the formal language, that he then tries to access in a measurement---as if Wigner had been his own friend.
In order to obtain a formal contradiction it is, in both cases, crucial to \emph{associate an account of experience}~(i.e., the statement to have observed a certain value) \emph{with symbols in a formal language}.
If the existence of such a formal account of experience is in doubt, then the formal contradiction is, too.

In this article, we argue against the \emph{exhaustive representation} of accounts of experience in any specific language, and, in particular, against the possibility to faithfully reduce them to symbols in a formal language.
We examine consequences for \emph{doing physics} in general, and for quantum mechanics and its measurement problem in particular.
Quantum mechanics may be complete insofar as it provides \emph{a description for any system}.
This description is, however, not necessarily exhaustive inasmuch as it contains anything that can be said about that system, or anything an observer associated with that system might say.

The outline of this article is as follows:
In Section~\ref{sec:theories_prop}, we characterize the epistemological role of the observer and isolate the crucial qualification of a measurement result as having been ``observed.''
In Section~\ref{sec:irred}, we review arguments against the possibility of a reduction of such a predicate to atomic rules of a formal language.
The resulting scepticism has consequences on how one regards the concepts of states, systems, and the role of the Born rule in quantum mechanics, as discussed in Section~\ref{sec:states_systems}.
In Section~\ref{sec:qm_observers}, we examine the measurement problem from the perspective developed in the preceding sections.

 \section{Theories, sentences, experience}
\label{sec:theories_prop}
\noindent
The starting point of our considerations are the following two characteristics which we regard as necessary~(albeit not sufficient) for doing \emph{physics}:
\begin{enumerate}[label={(C\arabic*)}]
  \item\label{c1} Physics strives for a \emph{formal description} of the world.
  \item\label{c2} \emph{Experience} serves as its final, normative authority.
\end{enumerate}
Concerning Characteristic~\ref{c2}: Within physics, one rarely refers to \emph{experience} but rather to \emph{observations}.
We take an observation to be a special kind of experience, e.g., verifiable in some sense.
Hence, any conclusion valid for experiences applies to observations.
In the following we attempt, to establish a notion of \emph{formal description} and to isolate the account of observation. 
In Section~\ref{sec:irred}, we address the question whether the program of formalization can be carried through to include the account of observation.
In contrast to, e.g.,~\cite{Fraassen1991}, we do not ``de-emphasize the role of language,'' but rather return to the case of the linguistic turn.

A \emph{formal language} is a set of sentences~$P\subset S^*$, with~$S^*$ being the Kleene closure~(the set of all finite strings or concatenations) of an alphabet~$S$, that are \emph{syntactically correct} with respect to a set of rules~$R$.\footnote{Tarski similarly characterizes ``formalized language'' in~\cite{Tarski36} and remarks the additional structure for ``formalized deductive sciences,'' where the rules are specified by axioms and deductive rules.}
The relation between observations and a formal language~$T=(S, R, P)$ is of the following kind~\cite{Popper1934}:
\begin{quote}
  \emph{If} I have observed~$x$, I deem a formal language~$T$ which forbade the observation of~$x$ untenable.
\end{quote}
If, for instance, in two subsequent identical quantum measurements, one observed different values, quantum mechanics,~$T^{\text{qm}}=(S^{\text{qm}}, R^{\text{qm}}, P^{\text{qm}})$, would be contradicted: 
There are no quantum states and measurement operators that could, within the postulates of quantum mechanics, account for the corresponding result~$(x_1,x_2)$ with~$x_1\neq x_2$.
The elements in~$S^{\text{qm}}$ together with rules in~$R^{\text{qm}}$ do not allow to conclude that~$x$ may be observed for either there is no such sentence in~$P^{\text{qm}}$ or an associated probability weight is zero.

\begin{figure}
\begin{center}
\begin{tikzpicture}[scale=.5]
  \def\h{1.5}
\def\v{2.5}
\def\lmu{0.05}
\coordinate (t) at (\h,\v);
\coordinate (tp) at (-\h,-\v);
\coordinate (ta) at (3.0*\h,-\v);
\draw[colorelement=red] (-1.5*\h,-1.5*\v) rectangle (t);
\draw[colorelement=magenta] (-\h,0) [sharp corners] -- (\h,0) -- (\h,-\v) [rounded corners] -- (tp) -- cycle;
\draw[colorelement=cyan] (\h, 0) [rounded corners]-- (3*\h,0) -- (ta) [sharp corners]-- (\h,-\v)  -- cycle;
\node[anchor=north east] at (t) {$P$};
\node[anchor=south east] at (ta) {$P^{\text{fal}}$};
\node[anchor=south west] at (tp) {$P'$};
\draw[thick, color=black, dashed, rounded corners] (-\h,0) rectangle (3*\h, -\v);
\node[anchor=south east] at (3*\h, 0) {$P^{\text{con}}$};

 \end{tikzpicture}
\end{center}
  \caption{Among all the syntactically correct sentences~$P^{\text{con}}$ about the experimental context, the sentences in~$P'$ are contained in the formal language~$T=(S,P,R)$, whereas the ones in~$P^{\text{fal}}$ contradict~$T$.}
\label{fig:fals_sent}
\end{figure}
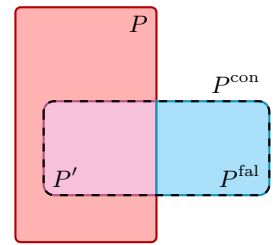

Let~$T=(S,R,P)$ be the formal language of some theory.
In order to formally characterize the sentences that falsify~$T$, we now consider statements restricted to a \emph{context}: 
We assume a formal language $T^{\text{con}}=(S^{\text{con}}, R^{\text{con}}, P^{\text{con}})$ whose syntactically correct sentences~$p\in P^{\text{con}}$ refer to \emph{possible results} observable for a given context, i.e., experimental setup.
Furthermore, we assume that the formal language $T$ can be restricted to some~$T'$ such that the corresponding sentences~$P'$ are the subset of $P^{\text{con}}$ of sentences that $T$ can account for.
For~$T$ to \emph{be falsifiable} in a context~$T^{\text{con}}$, the set of sentences~$P'$ has to be a proper subset of~$P^{\text{con}}$.
We can derive a formal language~$T^{\text{fal}}$ from~$T'$ and~$T^{\text{con}}$ with all sentences about the experimental context that contradict~$T$ as illustrated in Figure~\ref{fig:fals_sent}.
If we considered a sentence in~$P^{\text{fal}}$ true, then~$T$ would be falsified for the given context.
In the previous example of two identical subsequent quantum measurements, the result~$(x_1,x_2)$ with~$x_1\neq x_2$ corresponds to a sentence in~$P^{\text{fal}}$.

\emph{Falsification} refers to theories, including semantic concepts beyond a formal language, rather than to the formal language directly.\footnote{Popper in~\cite{Popper1934} refers to ``empirical-scientific systems.''}
The semantic concepts are delegated to the condition ``I have observed~$x$,'' which we now turn to.

Ultimately, sentences are to be gauged by experience according to Characteristic~\ref{c2}.
Let~$v_O:P^{\text{con}}\to\{ {\tt true}, {\tt false} \}$ be a function---the so-called \emph{verification function}---, for which~$v_O(p) = {\tt true}$ if and only if the observer~$O$ has observed~$p$.
More specifically, if~$v_O(p) = {\tt true}$ then~$O$ deems some apparatus fit to produce values in the context~$\mathcal{M}$, and~$O$ is certain to have obtained the value~$p$.
The verification function establishes whether the condition ``I have observed~$p$'' is satisfied or not.
Thus, for~$O$,~$T$ is falsified for the context~$P^{\text{con}}$ if there exists~$p\in P^{\text{fal}}$ with~$v_O(p) = {\tt true}$.

The reasoning about a formal language~$T$ is independent of the particular observer---it transcends single observers:
Any observer, upon knowing the formal language~$T$, can conclude for all sentences~$p\in P^{\text{con}}$ whether they are in~$P^{\text{fal}}$ or not and, hence, whether they falsify~$T$ or not.
The rules~$R$ are, therefore, required not to depend on observations, qualified by a---\emph{subjective}, as argued below---verification function.
A theory in this sense is \emph{not} statistical learning; thus, our approach differs from the one in~\cite{Mueller17}.\footnote{The symbols in~$S$ and the rules~$R$ allow to deduce all sentences in~$S$ that~$T$ accounts for independently of the prior observations such as a history in the \emph{observer graph} in~\cite{Mueller17}.}

Can the verification function be reduced to ``atomic''~(i.e., not further reducible) symbols and thus gain a similar observer-independent status?
Is there a \textcquote{Rorty1990}{single language sufficient to state all the truths there are to state}?\footnote{In~\cite{Rorty1990}, Rorty characterizes \emph{reductionism} as the attempt to find such a language.}
In Section~\ref{sec:irred}, we address these questions.

 \section{Reducible or not?}
\label{sec:irred}
\noindent
In the following, we argue that accounts of experience cannot be exhaustively expressed in a formal language.
In a first step, we examine how limits on formal languages restrict a formalizability of the verification function (see Section~\ref{ssec:limits_of_language}).
Even if one removes the restriction to \emph{formal} languages: There remain arguments against the existence of a \emph{privileged language} that reveals the one truth (see Sections~\ref{sub:semantic_holism} and~\ref{sub:no_privileged_language}).

In analogy to an observation by Putnam,\footnote{\blockcquote[\S5]{putnam1991representation}[.]{Reichenbach, Carnap, Hempel, and Sellars gave principled reasons why a finite translation of material-thing language into sense-datum language was impossible. 
Even if these reasons fall short of a strict mathematical impossibility proof, they are enormously convincing [\ldots]. 
In the same spirit, I am going to give principled reasons why a finite empirical definition of intentional relations and properties in terms of physical/computational relations and properties is impossible---reasons which fall short of a strict proof, but which are, I believe, nevertheless convincing}
 } it should be noted that the subsequent arguments do not constitute a \emph{proof} of impossibility.
We aim to show that the belief that there is \emph{one} language that unveils the unique truth, is merely one of multiple philosophical stances.\footnote{The arguments given here might not resonate easily:
If one adheres to the belief that there exists a privileged language accessing the truth, then probably one does not believe the statement that \emph{there is no privileged language} would be a true sentence of \emph{that} language. 
On the contrary: The latter statement is then easily dismissed.
This shows the \emph{contingency of language}---the incapability to step outside language, or into some meta-language that allows an innocent view on all languages~\cite{RortyCIS,WttgstnOgden90}.}
The readiness to associate symbols in a formal language with accounts of experience, as is necessary in formulating the measurement problem, stems from such a belief.
Though not logically excluded, such a belief is in tension with an essential idea of empirical sciences\footnote{See~\ref{c2}.}---and, as we argue below, an idea of how \emph{meaning} comes about: 
\emph{No theory can claim for itself to have exhaustively captured an observer's account of experience, while it draws legitimacy from experimental findings.}
Should not, instead, the role of an \emph{ironic}~\cite{RortyCIS} appeal to the (quantum) theorist, always doubtful towards one's \emph{final vocabulary}? 
 \subsection{The limits of formal languages}
\label{ssec:limits_of_language}
\noindent
\emph{Self-reference} imposes limitations on formal languages. 
Problems arising from self-reference crystallize, e.g., in the \emph{Liar's antinomy}: 
\begin{equation*}
  \text{This sentence is false.}
\end{equation*}
The sentence is self-contradictory because a statement is true if and only if the claim \emph{that the statement be true} is true.
This particular \emph{unquotation notion of truth}---illustrated by: ``Snow is white'' is true if and only if snow is white---is called \emph{Schema~T}~\cite{Tarski44}. 
For a truth predicate~$T$ and with $\langle\cdot\rangle$ denoting the name of a sentence, it can be written as
\begin{equation*}\label{eqn:schemaT}
  \phi \leftrightarrow T\langle\phi\rangle,\quad \forall \ \text{sentences} \ \phi \, .
\end{equation*}
With the theorem on the \emph{undefinability of truth}, Tarski showed that any formal language extending first-order arithmetic\footnote{A first-order arithmetic is an axiomatic system for the arithmetic of natural numbers relying merely on statements of first-order logic.} with a truth predicate containing Schema~T allows for such a contradiction.
Self-reference, together with a negation, 
thwarts the unification of different linguistic contexts as a consistent truth predicate requires a meta-language.

The verification function above serves as a predicate. 
If it is formalized in a theory extending first-order arithmetics, we require it to contain the Schema~T, and no sentences of the formal language are excluded from arising from observation, then there is a liar's antinomy.
There are three escape routes:
\begin{enumerate*}[label={(\arabic*)}]
  \item\label{la1} Either nature miraculously removes all problematic sentences, and leaves us with an \emph{incomplete} formal language in which certain syntactically correct sentences are excluded by assumption,
or 
  \item\label{la2} we can run into glitches in our experience that we cannot account for without contradiction, undermining our confidence in experience as the appropriate normative authority~(see also the reliability constraint in~\cite[2B]{GuptaEmpExp}),
or 
  \item\label{la3} we avoid the reference to sentences within that language.
\end{enumerate*}
Tarski follows the latter path: 
He confines the definition of truth to languages that are not semantically closed---languages that do not refer to their own sentences.
The definition of a truth predicate then becomes part of an expanded \emph{metalanguage}.
Following Tarski's path, we have to accept a dualism \emph{before} addressing the problematic dualism between states and the statistics of measurement results in quantum mechanics.
 \subsection{Semantic holism}
\label{sub:semantic_holism}
\noindent
The argument of \emph{semantic holism} has been turned against the existence of \emph{any} fundamental dualism.
Quine turns against the dogma of reductionism---\textcquote{Quine2D}{the belief that each meaningful statement is equivalent to some logical construct upon terms which refer to immediate experience}:
In a first step, he argues for a ``confirmation holism''---based on the observation \textcquote{Esfeld2001}{that experience can confirm or refute only a whole system of knowledge}.\footnote{The idea does not only originate from Quine; he rather develops Duhem's holism further~\cite{Esfeld2001}.}
By means of a confirmation theory of meaning, this entails semantic holism: \textcquote{Esfeld2001}[]{If a statement cannot be confirmed in isolation, it does not have meaning in isolation either}.
Gonseth observes that in light of semantic holism, no parts of language, not even logic, can be exempt from revision.\footnote{The principle of revision, one of four pillars of Gonseth's open philosophy, says \textcquote{Esfeld2001}[]{that every position and every scientific statement, including statements of logic, can be revised}.}

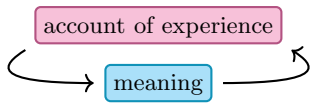
\begin{figure}
\begin{center}
\begin{tikzpicture}[scale=0.55]
  \node[colorbox=cyan] (m) at (0,-0.7) {meaning};
\node[colorbox=magenta] (aoe) at (0, .7) {account of experience};
\draw[conn=4pt, ->] (m.east) .. controls ($(m.east)+(2.8,.0)$) and ($(aoe.south east)+(.8,-.5)$) .. (aoe.south east);
\draw[conn=4pt, ->] (aoe.south west) .. controls ($(aoe.south west)+(-.8,-.5)$) and ($(m.west)+(-2.8,0)$) .. (m.west);
 \end{tikzpicture}
\end{center}
\caption{The circular dependency of meaning and the account of experience, eventually, calls into question any reduction to atoms.}
\label{fig:circ}
\end{figure}

Wittgenstein's \emph{Tractatus logico-philosophicus}~\cite{WttgstnOgden90}~(see also~\cite{SchulteWittgenstein}) shows similar holistic traits: If one joins the dots, a dense network of relations between terms emerges, but no \emph{ultimate explanation}.
Wittgenstein exposes the \textcquote{WttgstnOgden90}{misunderstanding of the logic of our language} indirectly.
He concludes:
\blockcquote[\S6.53]{WttgstnOgden90}{My propositions are elucidatory in this way: he who understands me finally recognizes them as senseless, when he has climbed out through them, on them, over them. (He must so to speak throw away the ladder, after he has climbed up on it.) 
    He must surmount these propositions; then he sees the world rightly.}
 So, how does meaning come about then?
In~\cite{WittgPhiloUntersuchungen}, after ``surmounting these propositions,'' Wittgenstein examines the root of first words considering language games (``Sprachspiele''):
For a child, that has no medium to establish meaning from explanation\footnote{\blockcquote[\S30]{WittgPhiloUntersuchungen}{Man muss schon etwas wissen (oder k\"{o}nnen), um nach der Benennung fragen zu k\"{o}nnen. Aber was muss man wissen?}}\label{fnWttg}, the acquisition of language reduces to trimming\footnote{\blockcquote[\S5]{WittgPhiloUntersuchungen}[.]{Das Lehren der Sprache ist hier [f\"{u}r das Kind] kein Erkl\"{a}ren, sondern ein Abrichten}
 }.
This process, in turn, relies on observation. 
{\em There is an issue of self-reference at the very root of meaning---or at the very root of any account of experience for that matter:}
Whenever one attempts to \emph{reduce} language to some \emph{atomic} propositions or first words, one is faced with a problem of circular dependence. 
Meaning and experience are intricately intertwined (see Figure~\ref{fig:circ}).

\subsection{No privileged language}
\label{sub:no_privileged_language}
\noindent
Wittgenstein's observations have been extended to argue against the existence of distinguished sentences that directly correspond to sensory input---atomic sentences whose meaning cannot be analyzed further:\footnote{Similar thoughts have been expressed by Nietzsche before: \blockcquote[p.18]{Nietzsche1873}[.]{Ein Maler, dem die H\"{a}nde fehlen und der durch Gesang das ihm vorschwebende Bild ausdr\"{u}cken wollte, wird immer noch mehr bei dieser Vertauschung der Sph\"{a}ren verrathen, als die empirische Welt vom Wesen der Dinge verr\"{a}th.
Selbst das Verh\"{a}ltnis  eines Nervenreizes zu dem hervorgebrachten Bilde ist an sich noch kein nothwendiges; [\ldots]
[D]as Hart- und Starr-Werden einer Metapher verb\"{u}rgt durchaus nichts f\"{u}r die Notwendigkeit und ausschliessliche Berechtigung dieser Metapher}
 }
Sellars argues, \textcquote[\S1]{Sellars1956}{as a first step in a general critique of the entire framework of givenness}:
Sense-datum theories are faced with an inconsistent triad, stemming from a concept of sensation as \textcquote[\S7]{Sellars1956}{inner episodes [\ldots] without any prior learning or concept formation} and \textcquote[\S7]{Sellars1956}{inner episodes which are non-inferential knowings that certain items are, for example, red or C{\#}~[the musical note] and that these episodes are the necessary conditions of empirical knowledge as providing the evidence for all other empirical propositions}.

Putnam argues against the existence of exhaustive criteria that determine \emph{reference} or \emph{representation}:
Does an ant's incidental ``picture of Churchill'' in the sand refer to Churchill? 
\emph{Similarity} to the features of Churchill is \emph{neither necessary nor sufficient} to refer to Churchill~(see~\cite[\S1]{putnam1981}).
Magritte's ``Treachery of Images'' exposes similar ambiguities:
Foucault distinguishes in his discussion of Magritte's references to a pipe~\cite{foucault1983} \emph{similitude} and \emph{resemblance}, and regards the painter to bring \textcquote[\S5]{foucault1983}[]{the former into play against the latter}.\footnote{\blockcquote[\S5]{foucault1983}[.]{Resemblance presupposes a primary reference that prescribes and classes. [\ldots] Resemblance serves representation, which rules over it; similitude serves repetition, which ranges across it}
 }
This leads to the question:
\textcquote[\S1]{putnam1981}[?]{If lines in the sand, noises, etc., cannot `in themselves', represent anything, then how is it that thought forms can `in themselves' represent anything}\footnote{\blockcquote[{\S}1.2]{RortyPMN}[.]{It seems perfectly clear, at least since Wittgenstein and Sellars, that the `meaning' of typographical inscription is not an extra `immaterial' property they have, but just their place in a context of surrounding events in a language-game, in a form of life.
This goes for brain-inscriptions as well}
 }

The assumption of innate, non-contextual semantic structures, an underlying \emph{lingua mentis} that allows to deduce any semantics if deciphered correctly is contested.
The existence of any one true and exhaustive language is in doubt.
And so are distinguished means to express the verification function~$v_O$, be they formal or not.
The argument is \emph{not} against a world out there, but against a \emph{truth out there}---a truth entirely beyond our creation~\cite{RortyCIS}.

 \section{Systems, States, and the Born rule}
\label{sec:states_systems}
\noindent
The hope of escaping the perils of practical activity and achieve ultimate certainty~\cite{DeweyQFC} is prevalent among quantum theorists.\footnote{The recent commotion about statements that suggest that quantum mechanics is inconsistent exemplify this attitude~\cite{FR18, BHW16}.}
This has effects on the use and meaning of terms.\footnote{\blockcquote[\S2]{putnam1991representation}[.]{As Wittgenstein often pointed out, a philosophical problem is typically generated in this way: certain assumptions are made which are taken for granted by \emph{all} sides in the subsequent discussion}
 } 
In the following, terms central to \emph{quantum mechanics} will briefly be discussed.

\subsection{State and system}
\noindent
As Wittgenstein observes, the meaning of terms is specified and modified while---not prior to---being used.
In this sense, the words ``state'' and ``system'' are not as clear and static as often insinuated in scientific practice:\footnote{Van Fraassen, in~\cite{Fraassen2006}, emphasizes the change of the use of words like ``atom'', ``electron'' or ``field'' to conclude: \textcquote{Fraassen2006}[]{Thus, scientific revolutions, \emph{and even evolutions}, embarrass the standard scientific realist}.}
This foundation for explanatory constructs might be shakier than it appears at first sight.

Whereas in classical mechanics, a \emph{system} and its \emph{state} are both characterized by directly measurable quantities, the matter gets more intricate in quantum mechanics:
\emph{Grete Hermann} refers to quantum states as ``new symbols that express the mutual dependency of the determinability of different measurements''\footnote{\blockcquote{hermann1935naturphilosophischen}[.]{Im Gegensatz dazu [Zust\"{a}nde der klassischen Physik] braucht der quantenmechanische Formalismus zur Zustandsbeschreibung neuartige Symbole, die die gegenseitige Abh\"{a}ngigkeit in der Bestimmbarkeit verschiedener Gr\"{o}{\ss}en zum Ausdruck bringt}
 }.
It seems reasonable to emphasize the \emph{semiotic} character of the state: 
Symbols of a formal language~$T$, though they might be called \emph{states}, are, a priori, not endowed with any further meaning than to serve as \emph{signs}.\footnote{\blockcquote[Bohr as quoted in][]{jammer1974}[.]{There is no quantum world. 
There is only an abstract quantum mechanical description. 
It is wrong to think that the task of physics is to find out how Nature is. 
Physics concerns what we can say about Nature}
 }
Gleason's theorem~\cite{Gleason57} sheds a new light onto the quantum state and shifts the semiotic character to measurements represented by sets of orthogonal projectors on a~(finite-dimensional) Hilbert space:
The theorem states that any measure~$\mu$ for projectors on a Hilbert space of dimension greater than two, satisfying an additivity constraint for orthogonal projectors,
\begin{equation*}
  \mu\left(\sum_i P_i\right) = \sum_i \mu(P_i)
\end{equation*}
as well as the constraints 
\begin{equation*}
  \mu( \bm 0) = 0 \qquad \mu(\bm 1) = 1
\end{equation*}
is of the form
\begin{equation*}
  \mu\left(P\right) = \tr( P \rho)
\end{equation*}
with~$\rho$ being a positive trace-1 operator, i.e., a density matrix.
A theory that represents measurements with sets of orthogonal projectors as above cannot bear a measure~$\mu$ with a range~$\{0,1\}$, as expected from a model assigning definite and non-contextual values to its measurement operators~\cite{redhead1989,sep-kochen-specker}. 
{\em The way back to a simple association of states and measurement results as in classical mechanics is barred by non-locality~\cite{hermann1935naturphilosophischen,Bell1964} and contextuality~\cite{KS67}.}

\subsection{Born rule}
\noindent
As demanded in Section~\ref{sec:theories_prop}, any physical theory has to yield assertions as to which results are \emph{not} expected to be observed.
In \emph{classical} mechanics, measurable entities have a one-to-one correspondence to states. 
The situation in quantum mechanics is more intricate, as there is no immediate correspondence between observed entities and state symbols:
The \emph{Born rule} bridges between these separate realms.
At first sight, the need of such a bridge seems unfortunate 
as it manifests the separation of a realm of observable entities from the realm of states.
Together with a model of the measurement process, the Born rule leads to issues such as the measurement problem.
On a closer look, however, fundamental theories seem to require a formal framework that connects to information:
The core of Popper's argument for a fundamental indeterminism in classical mechanics ($C$) in~\cite{PopperIndet1} evolves around the observation that the only means of extracting information from a system in classical mechanics---without going beyond classical mechanics in the measurement process---involves introducing untraceable disturbances.
The underlying assumption is: In order to extract any information, there has to be an interaction.\footnote{Bohr anticipated this observation: 
\blockcquote[]{Bohr1935}[]{Indeed the \emph{finite interaction between object and measuring agencies} conditioned by the very existence of the quantum of action entails---because of the impossibility of controlling the reaction of the object on the measuring instruments if these are to serve their purpose---the necessity of a final renunciation of the classical ideal of causality and a radical revision of our attitude towards the problem of physical reality}.}
The problem in classical mechanics is usually tamed by describing the measurement in \emph{another} theory~$T'$:
For instance, \emph{optical} measurement devices may determine the position of a system otherwise described by classical mechanics.
The disturbance due to the interaction captured in~$T'$ may be assumed to be small. 
Then the error one has to tolerate before falsifying~$C$ or~$T'$ is \emph{small}.
(In contrast to this, Popper's observation of a fundamental indeterminism relates to necessarily \emph{large} errors.)
In this light, the possibility of measuring an entity \textcquote{EPR}{without in any way disturbing a system} is \emph{not} a feature of classical mechanics \emph{alone}---\emph{optics} (or another external theory) is necessary to reduce the errors.

In summary, a \emph{fundamental} theory~(not relying on other theories to capture an interaction necessary to extract information) has to model the emergence of information---potentially exposing it to a measurement problem.
 \section{The measurement problem}
\label{sec:qm_observers}
\noindent
The above considerations have repercussions on the measurement problem. 
In the proceeding, we illustrate the problem with the \emph{Wigner's-friend experiment}.
A discussion of Maudlin's reading~\cite{Maudlin95} of the problem can be found in Appendix~\ref{ssec:maudlin}.
Breuer's approach~\cite{Breuer1995} to self-reference issues in quantum mechanics in is examined in Appendix~\ref{ssec:qm_self_ref}.

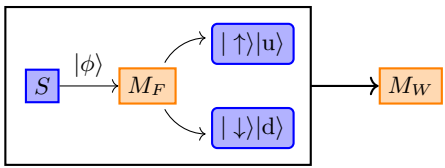
\begin{figure}
\begin{center}
  \begin{tikzpicture}[scale=0.7, font=\small]
  \draw[thick] (-.7,-1.5) rectangle (5.1,1.5);
  \node[fill=blue!30, draw=blue, thick] (s) at (0,0) {$S$};
  \node[fill=orange!30, draw=orange, thick] (m) at (2,0) {$M_F$};
  \node[fill=orange!30, draw=orange, thick] (M) at (7,0) {$M_W$};
  \node[fill=blue!30, draw=blue, thick, rounded corners=2pt] (r1) at (4,.8) {$\ket{\ua}\ket{\text{u}}$};
  \node[fill=blue!30, draw=blue, thick, rounded corners=2pt] (r2) at (4,-.8) {$\ket{\da}\ket{\text{d}}$};
  \draw[->] (s) -- node [midway,above] {$\ket{\phi}$} (m);
  \draw[->,shorten >=2pt,shorten <=2pt] (m) to[bend left] (r1);
  \draw[->,shorten >=2pt,shorten <=2pt] (m) to[bend right] (r2);
  \draw[->,thick] (5.1,0) to (M);
\end{tikzpicture}
 \end{center}
\caption{The Wigner's-friend experiment.}
\label{fig:wigners_friend}
\end{figure}

The measurement problem roots in the irreconcilability of distinguishability of information and linearity in quantum theory~\cite{BHW16}.
The problem can be illustrated with a Wigner's-friend experiment as depicted in Figure~\ref{fig:wigners_friend}.
Employing references to states---despite the above criticism---, the problem can be put as follows:
A source emits a system in a state 
\begin{equation*}\label{eqn:source_state}
  \ket{\phi}=\frac{\ket{0} + \ket{1}}{\sqrt{2}}\in\H_S\,.
\end{equation*}
This state is then measured by an observer, called \emph{Wigner's friend}, and modelled by a quantum system with Hilbert space~$\H_F$, in the basis~$\{ \ket{0}, \ket{1}\}$.
Finally, Wigner himself measures the joint system~$\H_S\otimes\H_F$ in a basis
\begin{IEEEeqnarray*}{RL}\label{eqn:meas_basis}
  \bigg\{ & \frac{\ket{0}_S\ket{0}_F+ \ket{1}_S\ket{1}_F}{\sqrt{2}}, 
   \frac{\ket{0}_S\ket{0}_F - \ket{1}_S\ket{1}_F}{\sqrt{2}}, \ \ldots \  \bigg\}\, . 
\end{IEEEeqnarray*}
Within Everett's \emph{relative-state formalism}~$Q$ of quantum mechanics~\cite{everett1957relative}\footnote{Here, we do not refer to the \emph{many-worlds interpretation} of quantum mechanics. We want to emphasize not to confuse the formalism with an interpretation~\cite{BW17}.}, the joint system~$\H_S\otimes \H_F$ after the friend's measurement is in a state
\begin{IEEEeqnarray*}{RL}\label{eqn:unitary_ref_state}
  V(\ket{\phi}_S) = & \frac{1}{\sqrt{2}} \big( V(\ket{0}_S) + V(\ket{1}_S) \big) \\
  = & \frac{1}{\sqrt{2}} \big( \ket{0}_S \otimes \ket{0}_F + \ket{1}_S\otimes\ket{1}_F\big)\, , 
\end{IEEEeqnarray*}
where~$V$ is the isometry modelling the measurement of the friend~\cite{BHW16}.
Thus, Wigner's final measurement yields the eigenvalue, corresponding to the first basis vector with probability 1.

Besides the relative-state formalism~$Q$, one might consider the scenario in a formalism~$Q_C$ with a \emph{``collapse.''} 
Within~$Q_C$, a system is either isolated and evolves unitarily according to the Schr\"{o}dinger equation, or it interacts with its environment. 
A \emph{measurement} is such an interaction and entails a collapse to the eigenvector associated with the measurement result.
In~$Q_C$, the measurement of the friend reduces to a unitary \emph{within} the system under consideration by Wigner:
\emph{Environment} is a subjective concept, depending on what an observer considers as the system to be measured. 
The system associated with~$\H_S\otimes\H_F$ measured by Wigner is \emph{isolated} from his perspective as long as he measures the entire joint system: The friend's ``measurement'' is \emph{not} an interaction with the environment relative to Wigner, but an interaction \emph{within} an isolated system.
Thus,~$Q_C$ yields predictions on the outcome of Wigner's final measurement identical to those of~$Q$.

In a third formalism with an \emph{objective collapse}~$Q_{OC}$, the joint system~$\H_S \otimes \H_F$ collapses, after the interaction modelled by the isometry~$V$ above conditioned on the obtained result. 
Importantly, this collapse is happening independently of the observer, not merely subjectively---that is, it has to be considered also by Wigner~(similarly to ``GRW''~\cite{GRW}).
For Wigner, who does not know the result of the friend's measurement, the entangled state~$V(\ket{\phi})$ collapses to a mixture 
\begin{equation}\label{eqn:mixed_state}
  \rho = \frac{1}{2} \left( \proj{0,0} + \proj{1,1} \right), 
\end{equation}
and Wigner measures either of two first vectors in the basis above with equal probability.

So, the question is: Does a measurement \emph{within} a closed system induce a collapse as modelled in~$Q_{OC}$?
The Wigner's-friend setup allows to empirically test this question and destinguish~$Q_{OC}$ from~$Q$/$Q_{C}$.

The precise meaning of the term ``measurement'' is crucial in the above considerations:
In~$Q$, any measurement is primarily a unitary, entangling the observed system with a memory system which includes the observer into the framework.
Even though a symbol in the formal language is assigned a tag ``the observer's memory,'' \emph{there is no simple correspondence to an ``observer'' in a discourse of ordinary language surrounding an experiment}. 
The Born rule is then applied to the memory system of the respective observer, after tracing out all other systems.
In~$Q_C$, the measurement is a notion \emph{relative} to the system that is to be measured.\footnote{This \emph{relativity} is reflected in the partial trace in~$Q$, which is relative to the memory of the respective observer, and in the unitary entangling the observed system with the observers memory.}
Any interaction that involves a part of the environment relative to this system can be considered a measurement and induces a collapse.
On the one hand, a system is called \emph{isolated} if there are no such interactions: 
Then, the evolution is a unitary operator.
As long as the perception of an ``isolated system'' on one side and a ``measurement'' or ``interaction with the environment'' on the other are clearly separated and mutually exclusive, this is a consistent framework. 
Note that the friend's measurement is \emph{not} a measurement relative to Wigner.

If~$Q$ and~$Q_C$ were falsified in a Wigner's-friend experiment, and we were left with~$Q_{OC}$, the situation would be intricate:
The relative character of the two other formalisms is lost, as there is an ``objective collapse.''
Then, an interaction with the environment or an associated dissipation of information is not a necessary requirement for a collapse.
{\em An isolated system would then show a non-unitary evolution.}
If the friend shared his result with Wigner, he would dissipate information into Wigner's environment, and, thus, we would again be in a case consistent with~$Q$ and~$Q_C$.
But can we establish that a measurement has happened without revealing the result?
One might assume that the friend's memory was in an initial state~$\ket{\Delta}$ orthogonal to both~$\ket{0}$ and~$\ket{1}$ and there was a unitary operator~$U$ with 
\begin{IEEEeqnarray*}{RrClCrClL}
  U: \ & \H_M & \otimes & \H_S & \to & \H_M & \otimes & \H_S&  \\
  &\ket{\Delta}& \otimes & \ket{k} & \mapsto & \ket{k} & \otimes & \ket{k} & \quad \forall k=1, \ldots, k . 
\end{IEEEeqnarray*}
for an orthonormal basis $\{\ket{k}\}_k$.
Then, a positive operator-valued measurement~(POVM) could determine whether the memory was still in the initial state or not without revealing the actual result.
If we were to build a consistent theory, then any other unitary of the same form as~$U$ should induce a collapse.
With this, quantum mechanics as we know it would eventually---\emph{collapse.}

In conclusion, the meaning of ``measurement'' is crucial to both~$Q$ and~$Q_C$: 
Different uses of the term---formal ones on the level of the theory, ordinary ones, different relative ones---\emph{are not to be simply equated}. 
The measurement as an account of observation---and thus of experience---does not have a direct correspondence to an interaction expressed in the formal language of a theory:
\emph{The measurement problem is a semantic confusion.}\footnote{Wigner already remarked on the imprecise use of terms: 
\textcquote[]{wigner1963problem}[]{Most importantly, he [the quantum theorist] has appropriated the word `measurement' and used it to characterize a special type of interaction by means of which information can be obtained on the state of a definite object. [\ldots] On the other hand, since he is unable to follow the path of the information until it enters his, or the observer's, mind, he considers the measurement completed as soon as a statistical relation has been established between the quantity to be measured and the state  of some idealized apparatus. He would do well to emphasize his rather specialized use of the word `measurement'\,}.}

 \section{Conclusion}
\noindent
Leibniz' reservation against space-time forming a fixed stage\footnote{The debate about a fundamental absolute space-time~\cite{vailati1997leibniz} appears in a new light if Reichenbach's principle is confronted with Bell non-locality, supporting the scepticism often attributed to Leibniz and not shared by many physicists throughout history---most notably Mach.} has its linguistic analogue:
Wittgenstein's observation on the intricate and circular dependency of meaning and experience lead him to conclude that meaning derives from the \emph{use} of language. 
These considerations on the plasticity of language have been extended by, e.g., Quine's arguments for semantic holism or Putnam's reflections on reference and representation, and lead, eventually, to doubting the existence of a privileged language---a ``semantic stage''---for accounts of experience:
\emph{Meaning arises along the discourse, not prior to it.}
There is, consequently, no objective language to exhaustively represent an observer's account of experience.

Reducing an ``observer'' to a ``system'' or a ``measurement'' as sensory perception to ``measurement'' as part of a formalism are light-footed extensions of the terms' use in physics.
In the face of the intricate relation between language and experience, as well as the normative authority of experience, it is debatable to \textcquote{Mueller17}[]{interchangeably use the words `experience', `observation,' and `state of the observer'\,}.
If an observer is measured, one can say \emph{something}---but not necessarily reproduce \emph{his} account of sensory evidence.

In summary: \emph{If \textbf{semantic} reductionism has limits, then so does \textbf{scientific} reductionism}.
It is in the context of the measurement problem and issues of self-reference that one runs into these limits.

In light of the absence of an underlying privileged language, \emph{physics is to \textbf{shape} meaning}.

\hbox{\ }
\begin{acknowledgments}
\noindent
  This work is supported by the Swiss National Science Foundation (SNF), the \emph{NCCR QSIT}, and the \emph{Hasler Foundation}. We would like to thank \"Amin Baumeler, Veronika Baumann, Cecilia Boschini, Paul Erker, Xavier Coiteux-Roy, Claus Beisbart, Manuel Gil, and Christian W\"{u}thrich for helpful discussions.
\end{acknowledgments}

\appendix
  \section{Maudlin's reading}
\label{ssec:maudlin}
\noindent
Maudlin in~\cite{Maudlin95} reads the measurement problem as the inconsistency of the following three ``claims:''

\begin{enumerate}[label={1.\Alph*}]
  \item\label{claim:compl} The wave-function of a system is \emph{complete}, i.e., the wave function specifies (directly or indirectly) all the physical properties of a system.
  \item\label{claim:lin} The wave-function always evolves in accord with a linear dynamical equation (e.g., the Schr\"{o}dinger equation).
  \item\label{claim:det} Measurements of, e.g., the spin of an electron always (or at least usually) have determinate outcomes, i.e., at the end of the measurement the measuring device is either in a state which indicates spin up (and not down) or down (and not up).
\end{enumerate}
To show the inconsistency, Maudlin considers, in addition to a two-dimensional observed system, a measurement apparatus. 
The argument is then: By linearity, the apparatus should be in a superposition of its pointer states ``up'' and ``down'' if the measured system is in a superposition with respect to the measurement basis (by~\eqref{claim:compl} and~\eqref{claim:lin}). 
According to~\eqref{claim:det}, on the contrary, the pointer of the apparatus should be exclusively in a state ``up'' or ``down.''

The commitment to the above ``claims'' can, however, be questioned as discussed below. 
Thus, for running into Maudlin's contradiction, one has to take a specific stance, namely \emph{assuming the sufficient representation of accounts by formal symbols possible}.
The resulting categorization of interpretations---by the choice of which claim to drop---is similarly subjective.

The completeness claim~\eqref{claim:compl} does not specify what is to be considered a ``physical property.''
Is it ``what can be measured''? Or ``what can be assigned a symbol in the formal language of a physical theory''?
It seems we are faced with similar problems as with what is to be considered a ``physical system.''
For the inconsistency to arise, accounts of experience are to be regarded as a ``physical property.''
The last claim,~\eqref{claim:det}, does not only state that measurements have determinate results, but it specifies these in terms of states of a measuring device.
{\em If one takes the requirement of ``determinate measurement results'' to be the capability to meaningfully assert ``I have observed `up' and not `down'\,'' (or vice versa)---that is, to attribute ``empirical content'' to such a statement---it does not necessarily follow that the state of some apparatus is exclusively in a correspondent state ``up'' or ``down.''}
One may include an apparatus into the formal framework and assume that then, the apparatus is being measured. 
Assuming the interaction between the observed system and the apparatus to be an entangling unitary means that the result of a measurement of \emph{the system} and the result of a measurement of \emph{the apparatus} have correlated results.

In a similar manner, further apparatuses ``between'' the observed system and the observer, who then reports on his experience, might be included.
One can even go as far as to assign a quantum state to the brain or memory of the observer---as we do above in connection with the relative-state formalism.
There are, however, doubts as to whether the assigned state---or any symbol in any other formal language---\emph{exhaustively} captures the meaning of an account as the one above.
If one follows, e.g., Quine's arguments on semantic holism, or Putnam's reflections on reference, such a \textcquote{Maudlin95}{representational completeness [of the wave-function]} seems out of reach:
There is then a gap for any formal language due to issues of semantics.
\emph{This semantic gap is wider than what a Born rule can bridge.}

The second claim,~\eqref{claim:lin}, states the linearity of quantum mechanics and refers to the Schr\"{o}dinger equation, as an example for such an evolution. 
The Schr\"{o}dinger equation does not hold unconditionally, but merely under the restriction to closed or isolated systems.
The claim that \textcquote{Maudlin95}{the traditional theory did \emph{not} [\ldots] state in clear physical terms, the conditions under which the non-linear evolution takes place} cannot be sustained if ``isolated systems'' is enough of a ``clear physical term.''
As argued above, the distinction between closed systems and systems interacting with their environment, e.g., during a measurement, is crucial in avoiding inconsistencies.

There is yet another intricacy:
If the system is isolated, then, according to the Schr\"{o}dinger equation, the system evolves \emph{unitarily}.
A system that interacts with its environment, i.e., that is \emph{not} isolated, evolves linearly according to some CPTP map.
This is the case for the formalism~$Q_{OC}$ above: 
There is a CPTP map so that
\begin{equation*}
  V(\ket{\phi}_S) \mapsto \rho\, ,
\end{equation*}
where~$\rho$ is the density matrix in~\eqref{eqn:mixed_state}.\footnote{The partial trace of an isometry~$V': \ket{l, l}_{SF} \mapsto \ket{l,l,l}_{SFG}$ modelling a second measurement of the friend in the same basis with a memory outside the control of Wigner, yields such a CPTP map.}
The divergence of the measurement probabilities for Wigner in~$Q_{OC}$ from the ones in~$Q$ and~$Q_{C}$ occurs already with a non-unitary, merely linear, evolution.

In light of the discussion in the previous chapters, a ``new physics'' that provides a \textcquote{Maudlin95}{real solution}, maybe even a \emph{realistic} solution, seems forlorn a hope.

   \section{Quantum self-reference}
\label{ssec:qm_self_ref}
\noindent
In~\cite{Breuer1995}, Breuer investigates problems of self-reference in quantum mechanics.
The approach relies on incorporating propositions and semantics into the formal language of quantum mechanics, and thus stands in contrast to the argumentation in Section~\ref{sec:irred} above.
Subsequently, the steps towards such an incorporation are examined in greater detail.

In a first step, the framework of propositions~$P$, corresponding to sentences about an experimental context we discuss in Section~\ref{sec:theories_prop}, is incorporated into~$T$. 
\blockcquote{Breuer1995}{Propositions about physical systems can be reformulated by saying  `The state of the system has this and that property.'
  So instead of speaking of propositions, we can equally well speak about sets of states: to each proposition there corresponds the set of states for which the proposition is true.}
 Propositions can be expressed in some formal language $T^{\text{con}}$ as done in Section~\ref{sec:theories_prop}. 
As such, there is little to object to incorporating~$T^{\text{con}}$ and~$T$ into some~$T'$.
The reduction of~$T^{\text{con}}$ \emph{into}~$T$ is, however, problematic. 
Let us take~$S$ to be the set of state symbols.
Then, the reduction proposed above can be regarded as a function~$f: P^{\text{con}} \to \mathcal{P}(S)$, assigning each proposition a set of states.
To speak ``equally well'' about sets of states, we require~$f$ to be injective.
The set of falsifying propositions,~$f^{-1}(\emptyset)$, has then a cardinality of one. 
This poses a caveat to falsifiability: 
If single sentences have no inherent meaning, then how can one meaningfully assert that the theory was contradicted?

One could allow for ordinary language to fill the void. 
This is not an option in the context of~\cite{Breuer1995}, as ordinary language is meant to be reduced to~$T$ as well.
\blockcquote{Breuer1995}[.]{So good experiments serve to at least partially constitute the semantics of physical theories. 
  In this sense, observation is a semantic concept}
 In light of the circular dependency of the account of experience and semantics discussed above, observation is a semantic concept.
An observation, however, is neither to be equated with an experiment nor with a formalism of ``measurement'' within~$T$, for reasons discussed in Section~\ref{sec:irred}.
Thus, the formal language of quantum mechanics is \emph{not} semantically closed, as concluded in the following statement.
\blockcquote{Breuer1995}[.]{Tarski (1956, 1969) calls a language semantically closed if it contains (1) semantic concepts and (2) expressions referring to its own propositions. 
The language of a physical theory can be closed semantically: If apparatus and object system, as well as their interaction, can be described by the theory, then the semantic concept of observation can be introduced into the language of the theory}
 In conclusion: One may doubt whether the reduction of the linguistic context, including the observer and his account of experience, to the formalism of quantum mechanics is feasible.
Consequently, an important premise for the arguments in~\cite{Breuer1995} is in question.

\end{document}